\documentclass[runningheads]{llncs}

\usepackage{graphicx}

\usepackage{caption}
\usepackage{subcaption}

\usepackage{tabularx}
\usepackage{ltablex}

\usepackage{listings}
\usepackage{color}
\usepackage{hyperref}
\usepackage{xcolor}

\newcommand\codetext[1]{\mbox{\texttt{#1}}}

\usepackage[strings]{underscore}

\begin{document}
\title{Refinement type contracts for verification of scientific investigative software}
%
%
\author{Maxwell Shinn\orcidID{0000-0002-7424-4230}}
\authorrunning{M. Shinn}
%
\institute{Yale University, New Haven CT 06520, USA}
\maketitle              
\begin{abstract}

  Our scientific knowledge is increasingly built on software output.
  User code which defines data analysis pipelines and computational
  models is essential for research in the natural and social sciences,
  but little is known about how to ensure its correctness.  The
  structure of this code and the development process used to build it
  limit the utility of traditional testing methodology.  Formal
  methods for software verification have seen great success in
  ensuring code correctness but generally require more specialized
  training, development time, and funding than is available in the
  natural and social sciences.  Here, we present a Python library
  which uses lightweight formal methods to provide correctness
  guarantees without the need for specialized knowledge or substantial
  time investment.  Our package provides runtime verification of
  function entry and exit condition contracts using refinement types.
  It allows checking hyperproperties within contracts and offers
  automated test case generation to supplement online checking.  We
  co-developed our tool with a medium-sized ($\approx$3000 LOC)
  software package which simulates decision-making in cognitive
  neuroscience.  In addition to helping us locate trivial bugs earlier
  on in the development cycle, our tool was able to locate four bugs
  which may have been difficult to find using traditional testing
  methods.  It was also able to find bugs in user code which did not
  contain contracts or refinement type annotations.  This demonstrates
  how formal methods can be used to verify the correctness of
  scientific software which is difficult to test with mainstream
  approaches.

\keywords{formal methods   \and scientific software \and contracts \and refinement types \and python \and runtime verification.}
\end{abstract}

\section{Introduction}

Over the last several decades, software engineering has made great
strides in developing tools and processes for verifying the
correctness of computer software.  Although verification has been
strikingly successful across many different domains, it has not been
widely applied to scientific software. It has been estimated that
5\%-100\% of scientific software output is incorrect due to undetected
software bugs \cite{Soergel2015}, and this is evident from the many
retractions caused by undetected bugs\footnote{See
  \url{http://retractionwatch.com/}}.
Some of these retractions were covered in depth by the popular
press. In one case, a researcher discovered that the results defining
his career were built on a software bug, forcing him to retract five
of his most important papers from top journals \cite{Miller2006}.  In
another case, a software bug reversed the results of highly
influential economics research which was widely cited in public policy
decisions \cite{Herndon2013}.  Section \ref{examples} will provide
concrete examples which illustrate common sources of such bugs.

One ubiquitous class of scientific software has received little to no
attention from the verification community: this software is user code
which is characterized by (a) small amounts of code (usually 50-500
LOC) (b) written by domain experts with little to no formal training
in software engineering \cite{Hannay2009,Chilana2009,Saltelli2014},
which is (c) run a limited number of times, (d) has specifications
that change unpredictably on a daily or hourly basis, (e) is used
almost exclusively by the original developer, and (f) has no testing
oracle because the output of the software is the object of
investigation \cite{Hatton1994,Kanewala2013,Soergel2015}.  We refer to
this class of software as ``investigative software''.  Investigative
software is written on a daily basis by countless researchers across
the natural and social sciences. Some common examples of investigative
software include: scripts to load experimental data and perform
statistical tests using statistical libraries; simulations of a
computational model; a pipeline which performs complicated
preprocessing operations on input data; or a script used internally to
make business decisions.  This paper presents a tool which is used to
verify investigative software.

Testing and verifying correctness of investigative software is
difficult due to many technical and cultural factors including unclean
data, the lack of a testing oracle, and the insufficiency of standard
testing procedures for the structure and goals of scientific programs
\cite{Kanewala2014,Hook2009,Johanson2018,Joppa2013,Kamali2015,Sanders2008}.
There already exist methods in the scientific community for checking
program correctness, but they have serious limitations which limit
their utility as discussed in Section \ref{otherwork}; as a result,
investigative software is most commonly validated by determining
whether the program's output matches the expectations of the
scientist, posing a fundamental violation of the scientific method
\cite{Sanders2008}.  A more convenient and effective method for
ensuring the quality of scientific software is needed.

Formal methods are able to check the correctness of investigative
software.  However, state of the art formal verification tools are
time-consuming to implement and require substantial formal methods
expertise, which make their use impractical for investigative
software.  A survey estimated that scientists spend about 30\% of
their time writing software \cite{Hannay2009}, so the need to formally
verify this code is a substantial hurdle to productivity.

By contrast, runtime verification is easy to use and especially
well-suited to investigative software.  It requires little training to
use, yet is able to achieve many of the same goals; because
investigative software is usually developed by its sole user, it
doesn't matter whether bugs are caught during development or detected
at runtime through verification condition violations.  This reflects
the fact that investigative software development often cannot be
meaningfully separated from data analysis
\cite{Paine2017,Hook2009,Kelly2015,Sanders2008}.  Researchers need
access to lightweight formal methods to improve the correctness of
their software without slowing down the research process.

We created Paranoid
Scientist\footnote{\url{https://github.com/mwshinn/paranoidscientist}},
a Python library for verifying the correctness of investigative
software.  Paranoid Scientist employs runtime verification to check
software correctness.  Developers specify function behavior through
entry and exit conditions, thereby creating function contracts which
must be satisfied by each function execution
\cite{Barnett2003,Hatcliff2012}.  
Contracts are specified in two parts.  First, each function argument
and the return value are specified modularly using refinement types
\cite{Freeman1994,Vazou2014}, 
which are types defined by predicates.  Second, additional constraints
such as a dependence of the return type on the function arguments or
the function argument types on each other are specified using predicates
written in pure Python.  These constraints may also depend on previous
calls to the function in order to check hyperproperties.  In addition
to runtime checking, Paranoid Scientist may further use the
refinement type contracts to automatically generate test cases.
Critically, Paranoid Scientist uses simple syntax and is intuitive to
those without a background in formal methods or software engineering.
We ran our tool on real-world investigative software and found that it
was able to catch four undetected bugs while imposing a 1.05--6.41
factor performance penalty.

\section{Motivating examples}\label{examples}

We provide two motivating examples from real-world investigative
software written for research in the biomedical sciences.

\subsubsection{Incorrect function usage}

Figure \ref{fig:matmult} shows an example bug in a function designed
to find the reverse complement of a DNA sequence.  Briefly, DNA is the
primary medium for long-term information storage in biological
organisms.  Each strand is composed of long sequences of four ``base''
molecules---adenine (\textsf{A}), guanine (\textsf{G}), cytosine
(\textsf{C}), and thymine (\textsf{T}).  For chemical stability, each
strand of DNA within an organism is usually bound to a ``reverse
complement'' strand whereby the \textsf{A} and \textsf{T} bases and
the \textsf{G} and \textsf{C} bases are swapped and the resulting
sequence is reversed.  By contrast, RNA is a biological medium for
short-term information storage.  It shares a nearly identical
structure with DNA, but replaces \textsf{T} with uracil (\textsf{U}).
Consequently, when forming the RNA reverse complement, \textsf{A} is
swapped with \textsf{U} instead of \textsf{T}.

The function \codetext{complement_sequence} in Figure
\ref{fig:matmult} uses a simple mechanism to compute the reverse
complement DNA sequence.  It accepts a list of characters and, using
Python's equivalent of the Unix program ``tr'', it converts all
\textsf{A} to \textsf{T}, \textsf{G} to \textsf{C}, \textsf{C} to
\textsf{G}, and \textsf{T} to \textsf{A}, and then reverses the result.
While this function behaves as intended, an erroneous usage of the
function occurs when the user tries to find the reverse complement of
an RNA sequence instead of a DNA sequence.  The function accepts the
valid RNA sequence \textsf{UCG} and returns another valid RNA sequence
\textsf{CGU} without error.  However, \textsf{CGU} is not the reverse
complement of \textsf{UCG}, because the \textsf{U} should be replaced
with an \textsf{A}. The \codetext{complement_sequence} function is
only designed to operate on DNA sequences, and thus cannot properly
deal with the RNA base \textsf{U}.

\begin{figure*}[t]
    \centering
    \begin{subfigure}[t]{0.52\textwidth}
\begin{lstlisting}[basicstyle=\scriptsize\ttfamily]
# Find the reverse  complement of 
# DNA. Assumes `seq` is a list of
# chars "A", "G", "C", and "T".
def complement_sequence(seq):
    # Convert `seq` to string and
    # use regex translate
    c_str = ''.join(seq).translate(
      str.maketrans('AGCT', 'TCGA'))
    return list(reversed(c_str))

# Correct usage, return ['G','A','T']
complement_sequence(['A','T','C'])
# BUG: Called for RNA instead of DNA,
# Return ['C','G','U'], however U 
# is not its own complement.
complement_sequence(['U','C','G'])
\end{lstlisting}
        \caption{Incorrect function usage example}
  \label{fig:matmult}
    \end{subfigure}%
    ~ 
    \begin{subfigure}[t]{0.51\textwidth}
\begin{lstlisting}[basicstyle=\scriptsize\ttfamily]
# Perform CPU-intensive preprocessing
timeseries = preprocess_ts(timeseries)
# Find timeseries pairwise correlations
corr_matrix = corrcoef(timeseries)
# Fisher z-transform (arctanh).
# Assumes abs(correlation) <= 1
norm = fisher_transform(corr_matrix)
save_csv("matrix.csv", norm)

corr_matrix = load_csv("matrix.csv")
# BUG: normalized before saving,
# these values violate assumptions
# of the fisher_transform function.
norm = fisher_transform(corr_matrix)
# Convert matrix to undirected graph
G = matrix_to_graph(norm)
\end{lstlisting}
\caption{NaN propagation example}
\label{fig:fmri}
    \end{subfigure}
    \caption{Two examples illustrate real-world bugs in investigative
      software.  (a) A function designed to process DNA sequences is
      invoked with an RNA sequence.  Due to their similarity, a valid
      but incorrect output is returned for the RNA sequence.  (b) NaN
      values can be silently propagated. This code constructs an
      undirected graph weighted by transformed pairwise correlations
      of brain scan timeseries.  The \codetext{fisher_transform}
      function is accidentally applied twice, the second execution of
      which generates a matrix with some NaN values which are silently
      ignored in the \codetext{matrix_to_graph} function.}
\end{figure*}

Paranoid Scientist is able to ensure correctness in this example.  An
annotated version of the \codetext{complement_sequence} function is
shown in Figure \ref{fig:matmultfixes}.  Most importantly for this
example, the \codetext{@accepts} decorator checks that the argument
value is a list consisting only of values \textsf{A}, \textsf{G},
\textsf{C}, and \textsf{T}.  This means that passing an RNA sequence
to the function in Figure \ref{fig:matmult} will raise an error,
because the ``U'' element in the list is invalid for DNA.  Even if
entry conditions were not specified, the ``U'' would have been caught
as an invalid output type by the return type specification in the
\codetext{@returns} decorator, and by the exit condition in the
\codetext{@ensures} decorator which specifies that all values of the
input must be different than the corresponding values of the reversed
output.  These properties are checked at runtime to ensure the
function is receiving correct input and producing correct output.

\begin{figure*}[h]
    \centering
    \begin{subfigure}[t]{0.56\textwidth}
\begin{lstlisting}[basicstyle=\scriptsize\ttfamily]
@accepts(List(Set('AGCT')))
@returns(List(Set('AGCT')))
@ensures("all(seq[i] != return[::-1][i] \
    for i in range(0, len(seq)))")
def complement_sequence(seq):
    ...
\end{lstlisting}
        \caption{Annotations for Figure \ref{fig:matmult}}
  \label{fig:matmultfixes}
    \end{subfigure}%
    ~ 
    \begin{subfigure}[t]{0.45\textwidth}
\begin{lstlisting}[basicstyle=\scriptsize\ttfamily]
@accepts(NDArray(t=Range(-1, 1)))
@returns(NDArray(t=Number))
@ensures("return.shape == \
    corr_values.shape")
def fisher_transform(corr_values):
    ...
\end{lstlisting}

        \caption{Annotations for Figure \ref{fig:fmri}}
\label{fig:fmrifixes}
    \end{subfigure}
    \caption{Two examples of Paranoid Scientist annotations which
      could detect the bugs in Figure \ref{fig:matmult} and
      \ref{fig:fmri}.  (a) Ensure that the \codetext{complement_sequence}
      function accepts and returns DNA sequences. (b) Ensure that the
      \codetext{fisher_transform} function only receives values for
      which it is defined, and that it always returns a number
      (i.e. not NaN or $\pm$inf).}
\end{figure*}

\subsubsection{NaN propagation} Another example bug is shown in Figure
\ref{fig:fmri}, which is a condensed version of a bug from real-world
investigative software.  This code is the final step in a pipeline
which converts a functional magnetic resonance imaging (fMRI) scan---a
type of brain scan which allows researchers to look at brain activity
over time---to an undirected graph where edges represent strong
correlations in brain activity \cite{Bullmore2009}.  First, several
computationally-intensive preprocessing steps are applied to the
timeseries, and then the pairwise Pearson correlation of each region
is computed.  Pearson correlation is a value from -1 to 1 inclusive,
which is converted to a more statistically-informative value from
negative infinity to infinity using the Fisher z-transform, or
equivalently hyperbolic arctangent.  This is then saved to a file so
that the computationally intensive steps do not need to be repeated.

After saving, the timeseries can be reloaded and turned into an
undirected graph.  However, before doing so, the
\codetext{fisher_transform} function is erroneously applied a second
time.  The second time it is called, the inputs hold values from
negative infinity to infinity.  This is outside of the domain of the
hyperbolic arctangent function, causing it to return NaN for inputs
less than -1 or greater than 1.  In practice, this gives reasonable
values for small- and medium-valued correlations but NaN for large
correlations.  These NaN values are masked by the function
\codetext{matrix_to_graph}, returning a graph which does not show
evidence of the NaN values; this creates a graph which appears to be
correct for all but the largest correlations.  This bug therefore
created subtle changes in the resulting graph topology which were not
noticed immediately, causing several weeks of work analyzing the
resulting graphs to be lost.

Paranoid Scientist is able to detect this bug. Figure
\ref{fig:fmrifixes} shows annotations for the
\codetext{fisher_transform} function.  This function takes any
$N$-dimensional array (\codetext{NDArray}) with values ranging from -1
to 1, and returns an $N$-dimensional array of the same shape with
elements which are numbers.  NaN is not a valid number.  These
annotations would have been sufficient to catch the bug in Figure
\ref{fig:fmri}.  Alternatively, annotations which specified the valid
input of \codetext{matrix_to_graph} would have also been able to
detect this bug.

If either of these two bugs had appeared in software which had an
oracle, there is a high probability that the difference in behavior
would manifest as an observable bug and the behavior could be
corrected \cite{Kanewala2013}.  Investigative software is written
because the result is unknown, so in these cases the bugs may never
have been found.

\section{Package summary}

\subsection{Refinement types}

Function entry and exit conditions are specified in part by refinement
types.  In our tool, refinement types are defined by a predicate which
checks whether an input is an element of the type.  Predicates are
constructed by Python functions using purely Python code, and thus may
reach arbitrary levels of complexity without depending on a domain
specific language.  This allows types to be defined in terms of their
scientific purpose and conceptual properties instead of as datatypes
\cite{Vazou2014}.  These refinement types are akin to what one would
write when documenting the function.  For example, one type defined by
default is ``Number'', which can be a float or an int but not NaN or
$\pm$inf.  Types can also represent more complex properties.  For
example, a discrete probability distribution is a list with
non-negative elements which sum to 1, and a correlation matrix is a
symmetric positive semidefinite matrix with 1 on the diagonal.  A list
of all types included by default is included in the Appendix.  Any
class can be used in place of a type by checking whether the passed
value is a subclass of the given class.  Alternatively, classes can
define a method to determine whether the passed value is an instance
of the class, described in further detail in Section \ref{syntax}.

\subsection{Entry and exit conditions}

In addition to refinement types, additional entry and exit conditions
may be specified for conditions involving multiple function arguments
or involving function arguments and the return value.  For example,
there may be a constraint that the first argument is greater than or
equal to the second argument, or that the function returns a matrix
with the number of rows and columns specified by the input arguments.
Conditions are specified as a string which is evaluated as Python
code.

Function properties may depend on more than a single execution of the
function.  For example, function concavity and function monotonicity
are hyperproperties which cannot be determined at runtime when
considering a single function execution.  Paranoid Scientist saves in
memory a list of arguments and return values from previous function
executions.  All future function executions are compared against these
past values.  For functions which are executed many times, a naive
implementation would cause serious performance and memory penalties,
limiting the practicality of this feature.  We address this problem by
saving only a subset of function calls and using reservoir sampling
\cite{Vitter1985} to test against a uniform distribution across all
function calls; as a result, verification of hyperproperties is not
performed across all previous calls but rather checked across a sample
of previous calls which is uniformly-distributed across time.  An
example of one such hyperproperty is shown in Figure \ref{fig:hyper}.

\begin{figure}[t]
\begin{lstlisting}[basicstyle=\scriptsize\ttfamily]
@accepts(Number)
@returns(Number)
@ensures("t >= t` --> return >= return`") # Monotonic function
def cube(t):
    return t**3
\end{lstlisting}
\caption{Example of a hyperproperty.  For any two executions of this
  function, Paranoid Scientist will check that the monotonicity
  hyperproperty is satisfied.}
\label{fig:hyper}
\end{figure}

\subsection{Syntax}\label{syntax}

Refinement types for function arguments and return values are
specified using the \codetext{@accepts} and \codetext{@returns}
function decorators, respectively.  Further entry and exit conditions
are passed as strings of Python code to the \codetext{@requires} and
\codetext{@ensures} function decorators, respectively.  The strings
are evaluated using a namespace which includes the function arguments
and additional user-specified libraries.  In \codetext{@ensures}, the
special value \codetext{return} represents the return value of the
function.  For testing hyperproperties, function arguments and return
values from previous executions can be accessed by appending one or
more backtick characters as a suffix, a notation which is reminiscent
of the ``prime'' symbol from mathematics.  Additionally, syntactic
sugar is available for the two common idioms ``implies'' with the
\codetext{-->} syntax and ``if and only if'' with the \codetext{<-->}
syntax.

In addition to the default types, refinement types may be defined
manually.  Types are classes which define two methods: one to test
values for adherence to the type, and a second to generate values of
the type for use in automated testing, as described below.  The type
definition may optionally accept arguments to specify parameterizable
behavior or generics.  Any existing class can also be used as a type
by using the class name in place of a refinement type.  In these
cases, Paranoid Scientist only tests whether the element is an
instance of the class, where adherence to the Liskov substitution
principle is assumed by default, i.e.\ subclasses are also considered
to be elements of their parent class.  For more precise control over
the checking, class methods may be defined analogous to the methods
for testing and generating values in stand-alone refinement types,
allowing a single class to serve both as a normal class and also as a
refinement type.

\subsection{Automated testing}

The use of entry and exit conditions for each function makes it
possible to perform unit tests automatically.  A stand-alone command
line utility takes the program to be checked as input and individually
tests each function in the program with values generated using a
specialized method from the refinement type specifications, similar to
fuzz testing \cite{Duran1984}.  These generated values are passed as
arguments to each function as long as the values satisfy the
function's entry conditions.  Because investigative software very
seldom includes tests, this increases robustness.

Not all functions can be tested. Those with unspecified types, strict
entry conditions, or arguments which cannot be automatically generated
will not produce any test cases.  Likewise, some tests run for a very
long time under certain parameterizations; these are killed after some
designated time duration to balance correctness with the practical
constraints of testing. Paranoid Scientist will report to the user a
list of the functions which could not be tested so that these may be
targeted for further testing.

More detailed information about syntax and automated testing is
available in the package documentation.

\section{Performance evaluation}

We evaluate the runtime performance of Paranoid Scientist on several
examples programs drawn from investigative software in cognitive
neuroscience.

\begin{description}
\item[Design matrix construction (\texttt{design})] Construct a design
  matrix for a generalized linear model, similar to the analysis
  performed in \cite{Park2014}.
\item[Nodal versatility (\texttt{versatility})] Compute the
  versatility \cite{Shinn2017} of a node in an undirected graph with
  respect to a community detection algorithm.
\item[Decision-making simulation (\texttt{pyddm_sim})] Simulate
  decision-making using the PyDDM software package (see Section
  \ref{sec:pyddm}).
\item[Decision-making fitting (\texttt{pyddm_fit})] Fit a
  decision-making model to simulated data using the PyDDM software
  package (see Section \ref{sec:pyddm}).
\end{description}

Performance benchmarks are shown in Table \ref{tab:performance}.
Overall, annotations comprised between 25-30\% of the lines of code.
There is a performance penalty for runtime verification, and this
penalty varies depending on the details of the code.  This penalty
falls within the ranges suggested for ``deliverable'' (a 3x slowdown)
or ``usable'' (10x slowdown or less) in runtime checks
\cite{Takikawa2016}.

\begin{table}[t]
  \centering
  \caption{\textbf{Performance benchmarks.} Runtime for each example
    program is shown with standard error of the mean over 10 runs.}
  {\small
    \begin{tabular}{l|llll}
     & \texttt{design} & \texttt{versatility} & \texttt{pyddm_sim} & \texttt{pyddm_fit} \\ \hline
    Total LOC & 162 & 86 & 5 & 20 \\
    Program LOC & 117 & 68 & 5 & 20\\
    Annotation LOC & 45 & 18 & (user code) & (user code)\\
    \% Code annotations & 28\% & 26\% & 0\% & 0\%\\
    Runtime w/ checking (s) & $8.452 \pm 0.052$ & $29.047 \pm 0.158$ & $13.685 \pm 0.057$ & $4.726 \pm 0.160$ \\
    Runtime w/o checking (s) & $3.239 \pm 0.022$ & $27.606 \pm 0.205$ & $2.136 \pm 0.013$ & $3.066 \pm 0.121$ \\
    Slowdown factor & 2.61 & 1.05 & 6.41 & 1.54 \\ 
  \end{tabular}
  \label{tab:performance} }
\end{table}

These examples were derived from real-world investigative software.
Notably, a previously undetected bug was found in the \texttt{design}
example when Paranoid Scientist annotations were added to it, which
caused incorrect binning of data before performing the regression
analysis.

\section{Case study}\label{sec:pyddm}

We used our tool while developing
PyDDM\footnote{\url{https://github.com/mwshinn/PyDDM}}, a
decision-making simulator for cognitive neuroscience.  PyDDM's
development was initially intended for a specific series of studies
and was later released to other research groups \cite{Gewaltig2014}.
Overall, Paranoid Scientist annotations comprised about 10\% of the
codebase.  Over 95\% of these annotations require refinement types
which cannot be checked using existing static type checkers for
Python.  Hyperproperties were specified for less than 5\% of function
exit conditions.

We briefly describe the motivation for PyDDM.  PyDDM aids in the study
of a simple form of decision-making whereby two options are presented
and the subject must choose one of the two based on either preference
or matching a given stimulus.  This type of decision-making is often
studied using the drift-diffusion model (DDM), which posits that all
decisions consist of some underlying evidence signal plus noise in
continuous time \cite{Ratcliff2008}.  The process of making a decision
relies on integrating evidence for each option over time and coming to
the final decision when the total integrated evidence surpasses some
confidence criterion.  Decision-making is usually studied for simple
decisions over a short duration of time (<5 sec), such as for
determining whether a sock is black or dark blue, but the model can
also be used for decisions which may span days or weeks, such as
deciding between two job offers.  In either case, there is a trade-off
between the time it takes to make the decision and the accuracy
expected.

The DDM represents evidence integration as a diffusion process
governed by the first passage time of a stochastic differential
equation across a boundary.  Analytical solutions for the DDM are fast
to compute, but only specific versions of the model can be solved
analytically.  Recent experiments have found that humans and animals
exhibit behavior which differs from these specific versions of the
model.  In order to explain experimental data, it is necessary to use
a more general version of the model which can only be solved
numerically.  PyDDM provides a consistent interface to a collection of
analytical and numerical algorithms for solving the generalized DDM,
selecting the best routine for each model.

As a result, PyDDM contains a mix of optimized routines for finding
solutions to stochastic differential equations, and a set of
object-oriented interfaces to make these routines convenient to use.
This includes predefined models for the most common use cases.
Consequently, PyDDM must be used in conjunction with user code written
by neuroscience researchers who may have limited experience in
software engineering.  Thus, the goals of verification are two-fold:
detecting errors in PyDDM itself, and detecting errors in PyDDM user
code.

\subsection{Detecting errors in PyDDM}

We found four non-trivial bugs in PyDDM using Paranoid Scientist.
Briefly, the bugs were:

\begin{enumerate}
\item Under certain circumstances, a small function which was assumed
  to return a positive number would return a negative number due to a
  typo in a mathematical equation.  This was caught by a return type
  of ``Positive''.
\item An algorithm assumed that the output of a previous step in the
  processing pipeline yielded a vector with 0 as the first element.
  It was intuitive that this should be the case.  When the numerical
  algorithm was upgraded, this no longer held true.  Only one distant
  branch in the pipeline relied on this value being zero.  This was
  caught by a precondition specifying the first element of the array
  should be 0.
\item When a model of a particular form was fit to data, a
  discretization approximation in a non-central portion of the code
  would exploit the limited numerical resolution of the simulation in
  order to select unnatural parameters which in turn artificially
  inflated the model's performance metrics.  This was caught by a
  postcondition which checked that the distribution would integrate to
  1.
\item Particular inputs caused numerical instabilities in one of the
  three simulation methods and made the probability distribution
  contain values slightly less than 0.  This was caught by a
  precondition on a different function which required all elements of
  the input array to be greater than 0.
\end{enumerate}

\noindent Bugs (1) and (2) would have been very difficult to detect
without our tool, and detecting bug (3) would have required manually
examining a large amount of intermediate output.  These three bugs
would have slightly impacted scientific results.  Bug (4) would have
likely been noticed eventually but would have caused a substantial
time investment to locate.  In addition to these bugs, Paranoid
Scientist was able to detect an internal inconsistency in how data
were stored.  Though this did not manifest in a bug which affected
results, it had the potential to do so in the future.

\subsection{Detecting errors using traditional methodology}

In addition to Paranoid Scientist annotations, unit tests and manual
code review were used to catch bugs in PyDDM.  One non-trivial bug was
found in unit testing which was not detected by Paranoid Scientist,
but this bug did not impact results:

\begin{enumerate}
\item When the core simulator's representation of a probability
  distribution was extended to include support for storing the
  distribution of incomplete trials, this unexpectedly modified the
  behavior of a distant piece of the code which utilized that
  representation.
\end{enumerate}

Additionally, two non-trivial bugs were detected through code review
which neither Paranoid Scientist nor unit/integration tests were able
to catch; these bugs also did not impact results:

\begin{enumerate}
\item When constructing a diffusion matrix as a part of the core
  simulation routine, certain rare but important cases utilized the
  previous timestep instead the next timestep.
\item A simulation algorithm is automatically chosen for each model,
  but this choice was suboptimal for a small number of models.
\end{enumerate}

\subsection{Detecting errors in user code}

In addition to PyDDM's core library code, a key feature is its
extensibility with user code to define new models.  Due to the
complexity of models which can be defined by users, it is important to
catch errors in user code even if the users do not use Paranoid
Scientist annotations.

Paranoid Scientist was able to find three bugs in user code, even
though this code did not have Paranoid Scientist annotations.  All
three of these would have impacted results:

\begin{enumerate}
\item Two subjects completed an experimental task with different task
  parameters, but parameters were mixed because the expression
  \codetext{if subject == 1} should have been \codetext{if subject !=
    1}.  This caused a parameterization which was valid on its own but
  not within the context of the data.  This was caught by a
  precondition which checked that one parameter was less than or equal
  to all elements of a data array.
\item A user-defined function to generate discrete probability
  distributions sometimes produced an invalid distribution.  This was
  caught by a precondition checking that the distribution summed to 1.
\item Boundaries were initialized randomly according to a normal
  distribution.  However, sometimes these bounds would be erroneously
  initialized such that some mass of the initial probability
  distribution had already crossed the bounds.  This was caught by a
  precondition which checked that two input vectors were the same
  size.
\end{enumerate}

\section{Limitations}

Runtime checking imposes penalties on the program's speed.  Paranoid
Scientist has not been optimized for speed, though such optimization
is possible in the future.  Previous work has demonstrated improved
performance of runtime checks through a client--server architecture
\cite{Dimopoulos2015} and through optional contracts
\cite{Dimoulas2013}.  Performance could also be improved by producing
a certificate during runtime which can be checked after execution.

Paranoid Scientist is compatible with all Python features and does not
require the programmer to limit herself to a Python subset or to use a
wrapper of the Python executable.  Nevertheless, some less-commonly
used Python features may cause problems if incorporated into contracts
due to the present implementation of runtime checking, especially if
these features are stateful.  For example, Python objects are allowed
to change their value when accessed, but this violates the assumptions
of runtime checking. Likewise, contracts cannot yet be specified for
generators.  If these features are needed in contracts, Paranoid
Scientist includes the \codetext{Unchecked} type as an alternative.

During runtime, Paranoid Scientist is able to implicitly deal with
side effects relevant to the results of the computation such as
modifications to global state and file IO.  However, automated tests
are unable to deal with these side effects.  Paranoid Scientist does
not have an explicit model of these or other side effects such as
exceptions and printing, because a clear specification of these is
seldom critical for investigative software.

Python's syntax for type
annotations\footnote{\url{https://www.python.org/dev/peps/pep-0484/}}
provides a convenient way to specify types.  Paranoid Scientist uses
function decorators instead of type annotations.  Type annotations
would be suitable for the \codetext{@accepts} and \codetext{@returns}
decorators, but not for the \codetext{@requires} or
\codetext{@ensures} decorators, so using decorators for all of these
improves syntax consistency.  The use of decorators allows type
annotations to be used for other purposes in the same codebase, and
may avoid confusion among less-experienced Python programmers who are
not used to this new syntax, or among users who run older versions of
Python.

\section{Related work}\label{otherwork}

\subsubsection{Formal methods for scientific software}

The present focus on investigative software differs from previous work
on verifying scientific software, which focuses on floating point
operations
\cite{Boldo2007,Goubault2011} 
or high performance computing \cite{Gunnels2001}. 
These tools are effective for specific types of scientific software,
but the methodology they impose does not reflect the environment in
which most investigative software is written and used
\cite{Sanders2008}.  Besides formal methods, prior work on testing
scientific software does not focus on investigative software; instead,
it focuses on large or collaborative software projects
\cite{Gewaltig2014,Carver2007,Sarma2016}, software written by seasoned
software engineers instead of researchers with limited formal training
\cite{Lundgren2016,Kanewala2013,Heaton2015}, software without an
oracle \cite{Kanewala2013}, or computationally- or
numerically-intensive software
\cite{Weyuker1982,Hochstein2008,Clune2011}.

\subsubsection{Testing scientific software}

Several recognized methods exist for software testing in the
scientific community, but these methods have serious limitations.  One
method involves rewriting a piece of software one or more times
by independent parties and comparing the output for identical input
\cite{Weyuker1982,Patel2017}.  In practice this is not feasible for
most investigative software, due to the fact that it is written by a
single individual for a limited number of executions.  Another method
is running the software with simplified parameters or artificial data
for which the result is known \cite{Weyuker1982,Patel2017}.  This
leaves the most scientifically important pieces of the code untested,
and it is often difficult to determine equivalence of the software's
output with the known result due to stochasticity or floating point
arithmetic \cite{Weyuker1982,Kanewala2014}. Meta-morphic testing has
been proposed as an alternative means of testing software without an
oracle, which involves testing specific properties which are required
to hold \cite{Patel2017,Chen2009,Giannoulatou2014,Chen1998}. 
This requires a deep knowledge of testing methodology and a code
structure which facilitates such tests.

\subsubsection{Python type systems}

In recent years there has been a proliferation of static type checkers
within the Python ecosystem, starting with
MyPy\footnote{\url{http://mypy-lang.org/}} and continued by Facebook's
Pyre\footnote{\url{https://pyre-check.org/}}, Google's
PyType\footnote{\url{https://google.github.io/pytype/}}, and
Microsoft's
PyRight\footnote{\url{https://github.com/Microsoft/pyright}}.  This
has been further advanced by PEP
484\footnote{\url{https://www.python.org/dev/peps/pep-0484/}}, which
standardized a syntax for type annotations for functions in the Python
language.  These type checkers introduce neither overhead nor
speedups, as the types are checked before and not during runtime.
Thus, the type information is not enforced during program execution.

Reticulated Python \cite{Vitousek2014} includes runtime checks for
annotated types using three different methods.  The two bugs described
in \cite{Vitousek2014} which were caught using Reticulated Python
would have occurred as exceptions without the runtime checks; with the
present work, we are more interested in bugs which might have
otherwise gone undetected.  One of Reticulated Python's modes of
operation can insert undetected bugs into the code by not preserving
object identity, demonstrating the different objectives between
Reticulated Python and the present work.  Additional packages for
runtime checking of static data types in Python, such as the
``enforce''\footnote{\url{https://github.com/RussBaz/enforce/}} or
``typeguard''\footnote{\url{https://github.com/agronholm/typeguard}}
packages, share many similarities to the ``transient'' method in
Reticulated Python.

\subsubsection{Python contracts libraries}

While contracts were first conceived for Python in the language's
infancy \cite{Ploscha1997}, contracts are still rare in Python code.
The most popular contract library for Python is
``PyContracts''\footnote{\url{https://andreacensi.github.io/contracts/}},
which embeds a domain specific language into Python for specifying
properties that each argument must satisfy.  It is difficult to
specify complex properties or to create properties which rely on more
than one argument, such as ``argument 1 is greater than argument 2''.
PyBlame \cite{Arai2016} provides a sophisticated contract library for
Python which integrates with the debugger, but a detailed comparison
could not be performed due to the lack of availability of the PyBlame
source code.  Data validation libraries, such as
``cerberus''\footnote{\url{https://docs.python-cerberus.org/en/stable/}}
and
``voluptuous''\footnote{\url{https://github.com/alecthomas/voluptuous}},
ensure that datasets satisfy particular conditions, and thus may be
used in conjunction with Paranoid Scientist.

Nagini is a package which provides full static verification for a
Python subset \cite{Eilers2018}.  Python code is converted to an
intermediate language and conditions are specified using contracts.
In addition to arbitrary assertions, it can reason about exceptions,
memory safety, data-race conditions, and input--output.  However,
approximately half of the lines of code must be devoted to the
specification, and it has difficulty inferring properties about
non-Python code such as C libraries.

\section{Conclusions and future directions}

It is difficult to overestimate the importance of investigative
software in scientific research, but few studies have examined
effective techniques for ensuring its correctness.  Paranoid Scientist
uses lightweight formal methods to provide correctness guarantees
about this difficult-to-test class of software.  It does so through a
combination of contracts and refinement types in a way which is easy
to use for those without explicit training in formal methods.  We
demonstrated that Paranoid Scientist can be used to find bugs in
scientific software which would have impacted results and would have
been otherwise difficult to detect.

Investigative software is built in an environment which poses two
unique challenges for Hoare-style verification and static checking of
preconditions and postconditions.  First, the verification technique
must be usable by scientists with little to no training in computer
science. The state of the art techniques require a deep knowledge of
formal methods to use them effectively \cite{Eilers2018}.  Second, the
amount of time spent verifying the software must be small compared to
the amount of time spent writing the code to be verified.  Current
techniques are time consuming to implement.  By contrast, our tool
requires approximately as much time to implement as does writing
function documentation.  

Technical constraints of investigative software raise further
challenges for formal methods.  Investigative software in Python
relies heavily on non-Python code such as C libraries and integrated
shell commands, and thus static verification would require scaffolding
for usability in practice.  This scaffolding mandates more effort for
verification and strong familiarity of the user with formal methods,
exacerbating the previously discussed environmental challenges.
Additionally, techniques such as type inference are conceptually
incompatible with the present approach because the types in Paranoid
Scientist often depend on the purpose of the code; for example, it may
be ``valid'' to accept a probability less than 0 but this does not
make sense on a scientific level.  A gradual static verification
approach, analogous to gradual typing \cite{Sieka2007}, may be useful.

Lightweight formal methods may also be applied to investigative
software in other programming languages.  The present work targets
Python due to its ubiquity in investigative software and powerful
metaprogramming capabilities to simplify implementation.  Besides
Python, common languages for investigative software include Matlab,
Julia, and R.  While the bug shown in Figure \ref{fig:fmri} may not
have occurred in other languages, different programming languages have
different advantages and disadvantages for the correctness of
investigative software.  For example, Matlab by default defines the
constants \codetext{i} and \codetext{j} to be $\sqrt{-1}$, but allows
these to be assigned other values by users.  As a result, mistaken
variable initialization or names can cause undetected bugs.
Additionally, unexpected files saved to locations in Matlab's
\texttt{PATH} can cause erroneous versions of scripts or data to be
loaded, or even cause built-in functions to change their behavior.  An
implementation of lightweight formal methods as described here would
be able to catch these and other bugs in Matlab.

It is critical to verify the correctness of investigative software,
but technical and cultural constraints limit the effectiveness of
conventional techniques.  Lightweight formal methods as implemented in
Paranoid Scientist provide a convenient and effective way to check the
correctness of investigative software.

\section*{Acknowledgments}

Thank you to Ruzica Piskac and Anastasia Ershova for a critical review
of the manuscript; Clarence Lehman, Daeyeol Lee, and John Murray for
helpful discussions; Michael Scudder for PyDDM code reviews; and
Norman Lam for PyDDM development and code reviews. Funding was
provided by the Gruber Foundation.

\appendix
\setcounter{secnumdepth}{0}
\section*{Appendix: Default types}

\scriptsize
\begin{tabularx}{\linewidth}{lX}
    \multicolumn{2}{l}{\textbf{\footnotesize Numerical}} \\
Numeric & A floating point or integer \\
ExtendedReal & A floating point or integer, excluding NaN \\
Number & A floating point or integer, excluding NaN and $\pm$inf \\
Integer & An integer \\
Natural0 & An integer greater than or equal to zero \\
Natural1 & An integer greater than zero \\
Range & A number with a value between two specified numbers, inclusive \\
RangeClosedOpen & A number with a value between two specified  numbers, inclusive on the bottom and exclusive on the top \\
RangeOpenClosed & A number with a value between two specified  numbers, exclusive on the bottom and inclusive on the top \\
RangeOpen & A number with a value between two specified  numbers, exclusive \\
Positive0 & A number greater than or equal to zero \\
Positive & A number greater than zero \\
NDArray & A Numpy \codetext{ndarray}, optionally with a given   dimensionality or elements which satisfy a given type  \vspace{.2cm}\\

  \multicolumn{2}{l}{\textbf{\normalsize Strings}} \\
String & A Python string \\
Identifier & A non-empty alphanumeric string with underscores and hyphens \\
Alphanumeric & A non-empty alphanumeric string \\
Latin & A non-empty string with Latin characters only \vspace{.2cm} \\

  \multicolumn{2}{l}{\textbf{\footnotesize Collections}} \\
Tuple & A Python tuple, with elements which satisfy given types \\
List & A Python list, with elements which satisfy a given type \\
Dict & A Python dictionary, keys and values which satisfy given types \\
Set & A Python set, with elements which satisfy a given type \\
ParametersDict & A dictionary which may include only a subset of keys, with values which satisfy given types \vspace{.2cm} \\

  \multicolumn{2}{l}{\textbf{\footnotesize Logical types}} \\
And & Logical AND of two or more types \\
Or & Logical OR of two or more types \\
Not & Logical NOT of a type \vspace{.2cm}\\

  \multicolumn{2}{l}{\textbf{\footnotesize Special types}} \\
Boolean & Either \codetext{True} or \codetext{False} \\
Function & A Python function  \\
Constant & A single specified value is accepted \\
Nothing & Only \codetext{None}, equivalent to \codetext{Constant(None)} \\
Unchecked & Any value (always succeeds) \\
Void & No value is accepted (always fails) \\
Maybe & Either a value of the specified type or else \codetext{None} \\
Self & The \codetext{self} argument to a method \\
PositionalArguments & Optional positional arguments to functions \\
KeywordArguments & Optional keyword arguments to functions \\

\end{tabularx}

\bibliographystyle{splncs04}
\bibliography{paranoid-submitted-refs}

\begin{thebibliography}{10}
\providecommand{\url}[1]{\texttt{#1}}
\providecommand{\urlprefix}{URL }
\providecommand{\doi}[1]{https://doi.org/#1}

\bibitem{Arai2016}
Arai, R., Sato, S., Iwasaki, H.: A debugger-cooperative higher-order contract
  system in {Python}. In: Asian Symposium on Programming Languages and Systems.
  pp. 148--168. Springer (2016)

\bibitem{Barnett2003}
Barnett, M., Schulte, W.: Runtime verification of .{NET} contracts. Journal of
  Systems and Software  \textbf{65}(3),  199--208 (Mar 2003)

\bibitem{Boldo2007}
Boldo, S., Filliatre, J.C.: Formal verification of floating-point programs. In:
  18th {IEEE} Symposium on Computer Arithmetic ({ARITH} '07). {IEEE} (Jun 2007)

\bibitem{Bullmore2009}
Bullmore, E., Sporns, O.: Complex brain networks: graph theoretical analysis of
  structural and functional systems. Nature Reviews Neuroscience
  \textbf{10}(3),  186--198 (Feb 2009)

\bibitem{Carver2007}
Carver, J.C., Kendall, R.P., Squires, S.E., Post, D.E.: Software development
  environments for scientific and engineering software: A series of case
  studies. In: 29th International Conference on Software Engineering
  ({ICSE}'07). {IEEE} (May 2007)

\bibitem{Chen2009}
Chen, T., Ho, J.W., Liu, H., Xie, X.: An innovative approach for testing
  bioinformatics programs using metamorphic testing. {BMC} Bioinformatics
  \textbf{10}(1), ~24 (2009)

\bibitem{Chen1998}
Chen, T.Y., Cheung, S.C., Yiu, S.M.: Metamorphic testing: a new approach for
  generating next test cases. techreport HKUST-CS98-01, The Hong Kong
  University of Science and Technology (1998)

\bibitem{Chilana2009}
Chilana, P.K., Palmer, C.L., Ko, A.J.: Comparing bioinformatics software
  development by computer scientists and biologists: An exploratory study. In:
  2009 {ICSE} Workshop on Software Engineering for Computational Science and
  Engineering. {IEEE} (May 2009)

\bibitem{Clune2011}
Clune, T.L., Rood, R.B.: Software testing and verification in climate model
  development. {IEEE} Software  \textbf{28}(6),  49--55 (Nov 2011)

\bibitem{Dimopoulos2015}
Dimopoulos, S., Krintz, C., Wolski, R., Gupta, A.: {SuperContra}:
  Cross-language, cross-runtime contracts as a service. In: 2015 {IEEE}
  International Conference on Cloud Engineering. {IEEE} (Mar 2015)

\bibitem{Dimoulas2013}
Dimoulas, C., Findler, R.B., Felleisen, M.: Option contracts. In: Proceedings
  of the 2013 {ACM} {SIGPLAN} international conference on Object oriented
  programming systems languages {\&} applications - {OOPSLA} '13. {ACM} Press
  (2013)

\bibitem{Duran1984}
Duran, J.W., Ntafos, S.C.: An evaluation of random testing. {IEEE} Transactions
  on Software Engineering  \textbf{{SE}-10}(4),  438--444 (Jul 1984)

\bibitem{Eilers2018}
Eilers, M., Müller, P.: Nagini: A static verifier for {Python}. In: Computer
  Aided Verification, pp. 596--603. Springer International Publishing (2018)

\bibitem{Freeman1994}
Freeman, T.: Refinement Types for ML. Ph.D. thesis, Carnegie Mellon University,
  Pittsburgh, PA, USA (1994)

\bibitem{Gewaltig2014}
Gewaltig, M.O., Cannon, R.: Current practice in software development for
  computational neuroscience and how to improve it. {PLoS} Computational
  Biology  \textbf{10}(1),  e1003376 (Jan 2014)

\bibitem{Giannoulatou2014}
Giannoulatou, E., Park, S.H., Humphreys, D.T., Ho, J.W.: Verification and
  validation of bioinformatics software without a gold standard: a case study
  of {BWA} and bowtie. {BMC} Bioinformatics  \textbf{15}(Suppl 16), ~S15 (2014)

\bibitem{Goubault2011}
Goubault, E., Putot, S.: Static analysis of finite precision computations. In:
  Lecture Notes in Computer Science, pp. 232--247. Springer Berlin Heidelberg
  (2011)

\bibitem{Gunnels2001}
Gunnels, J.A., van~de Geijn, R.A.: Formal methods for high-performance linear
  algebra libraries. In: {IFIP} Advances in Information and Communication
  Technology, pp. 193--210. Springer {US} (2001)

\bibitem{Hannay2009}
Hannay, J.E., MacLeod, C., Singer, J., Langtangen, H.P., Pfahl, D., Wilson, G.:
  How do scientists develop and use scientific software? In: 2009 {ICSE}
  Workshop on Software Engineering for Computational Science and Engineering.
  {IEEE} (May 2009)

\bibitem{Hatcliff2012}
Hatcliff, J., Leavens, G.T., Leino, K.R.M., Müller, P., Parkinson, M.:
  Behavioral interface specification languages. {ACM} Computing Surveys
  \textbf{44}(3),  1--58 (Jun 2012)

\bibitem{Hatton1994}
Hatton, L., Roberts, A.: How accurate is scientific software? {IEEE}
  Transactions on Software Engineering  \textbf{20}(10),  785--797 (1994)

\bibitem{Heaton2015}
Heaton, D., Carver, J.C.: Claims about the use of software engineering
  practices in science: A systematic literature review. Information and
  Software Technology  \textbf{67},  207--219 (Nov 2015)

\bibitem{Herndon2013}
Herndon, T., Ash, M., Pollin, R.: Does high public debt consistently stifle
  economic growth? {A} critique of {Reinhart} and {Rogoff}. Cambridge Journal
  of Economics  \textbf{38}(2),  257--279 (Dec 2013)

\bibitem{Hochstein2008}
Hochstein, L., Basili, V.: The {ASC}-alliance projects: A case study of
  large-scale parallel scientific code development. Computer  \textbf{41}(3),
  50--58 (Mar 2008)

\bibitem{Hook2009}
Hook, D., Kelly, D.: Testing for trustworthiness in scientific software. In:
  2009 {ICSE} Workshop on Software Engineering for Computational Science and
  Engineering. {IEEE} (May 2009)

\bibitem{Johanson2018}
Johanson, A., Hasselbring, W.: Software engineering for computational science:
  Past, present, future. Computing in Science {\&} Engineering pp.~1--1 (2018)

\bibitem{Joppa2013}
Joppa, L.N., McInerny, G., Harper, R., Salido, L., Takeda, K., O'Hara, K.,
  Gavaghan, D., Emmott, S.: Troubling trends in scientific software use.
  Science  \textbf{340}(6134),  814--815 (May 2013)

\bibitem{Kamali2015}
Kamali, A.H., Giannoulatou, E., Chen, T.Y., Charleston, M.A., McEwan, A.L., Ho,
  J.W.K.: How to test bioinformatics software? Biophysical Reviews
  \textbf{7}(3),  343--352 (Aug 2015)

\bibitem{Kanewala2013}
Kanewala, U., Bieman, J.M.: Techniques for testing scientific programs without
  an oracle. In: 2013 5th International Workshop on Software Engineering for
  Computational Science and Engineering ({SE}-{CSE}). {IEEE} (May 2013)

\bibitem{Kanewala2014}
Kanewala, U., Bieman, J.M.: Testing scientific software: A systematic
  literature review. Information and Software Technology  \textbf{56}(10),
  1219--1232 (Oct 2014)

\bibitem{Kelly2015}
Kelly, D.: Scientific software development viewed as knowledge acquisition:
  Towards understanding the development of risk-averse scientific software.
  Journal of Systems and Software  \textbf{109},  50--61 (Nov 2015)

\bibitem{Lundgren2016}
Lundgren, A., Kanewala, U.: Experiences of testing bioinformatics programs for
  detecting subtle faults. In: Proceedings of the International Workshop on
  Software Engineering for Science - {SE}4Science '16. {ACM} Press (2016)

\bibitem{Miller2006}
Miller, G.: A scientist's nightmare: Software problem leads to five
  retractions. Science  \textbf{314}(5807),  1856--1857 (Dec 2006)

\bibitem{Paine2017}
Paine, D., Lee, C.P.: "{Who} has plots?". Proceedings of the {ACM} on
  Human-Computer Interaction  \textbf{1}({CSCW}),  1--21 (Dec 2017)

\bibitem{Park2014}
Park, I.M., Meister, M.L.R., Huk, A.C., Pillow, J.W.: Encoding and decoding in
  parietal cortex during sensorimotor decision-making. Nature Neuroscience
  \textbf{17}(10),  1395--1403 (Aug 2014)

\bibitem{Patel2017}
Patel, K., Hierons, R.M.: A mapping study on testing non-testable systems.
  Software Quality Journal  \textbf{26}(4),  1373--1413 (Nov 2017)

\bibitem{Ploscha1997}
Plosch, R.: Design by contract for {Python}. In: Proceedings of Joint 4th
  International Computer Science Conference and 4th Asia Pacific Software
  Engineering Conference. {IEEE} Comput. Soc (1997)

\bibitem{Ratcliff2008}
Ratcliff, R., McKoon, G.: The diffusion decision model: Theory and data for
  two-choice decision tasks. Neural Computation  \textbf{20}(4),  873--922 (Apr
  2008)

\bibitem{Saltelli2014}
Saltelli, A., Funtowicz, S.: When all models are wrong. Issues in Science and
  Technology  \textbf{30}(2),  79--85 (2014)

\bibitem{Sanders2008}
Sanders, R., Kelly, D.: Dealing with risk in scientific software development.
  {IEEE} Software  \textbf{25}(4),  21--28 (Jul 2008)

\bibitem{Sarma2016}
Sarma, G.P., Jacobs, T.W., Watts, M.D., Ghayoomie, S.V., Larson, S.D., Gerkin,
  R.C.: Unit testing, model validation, and biological simulation.
  F1000Research  \textbf{5}, ~1946 (Aug 2016)

\bibitem{Shinn2017}
Shinn, M., Romero-Garcia, R., Seidlitz, J., V{\'{a}}{\v{s}}a, F., V{\'{e}}rtes,
  P.E., Bullmore, E.: Versatility of nodal affiliation to communities.
  Scientific Reports  \textbf{7}(1) (Jun 2017)

\bibitem{Sieka2007}
Siek, J., Taha, W.: Gradual typing for objects. In: European Conference on
  Object-Oriented Programming, pp. 2--27. Springer Berlin Heidelberg (2007)

\bibitem{Soergel2015}
Soergel, D.A.W.: Rampant software errors may undermine scientific results.
  F1000Research  \textbf{3}, ~303 (Jul 2015)

\bibitem{Takikawa2016}
Takikawa, A., Feltey, D., Greenman, B., New, M.S., Vitek, J., Felleisen, M.: Is
  sound gradual typing dead? In: Proceedings of the 43rd Annual {ACM}
  {SIGPLAN}-{SIGACT} Symposium on Principles of Programming Languages - {POPL}
  2016. {ACM} Press (2016)

\bibitem{Vazou2014}
Vazou, N., Seidel, E.L., Jhala, R., Vytiniotis, D., Peyton-Jones, S.:
  Refinement types for {Haskell}. In: Proceedings of the 19th {ACM} {SIGPLAN}
  international conference on Functional programming - {ICFP} '14. {ACM} Press
  (2014)

\bibitem{Vitousek2014}
Vitousek, M.M., Kent, A.M., Siek, J.G., Baker, J.: Design and evaluation of
  gradual typing for {Python}. In: Proceedings of the 10th {ACM} Symposium on
  Dynamic languages - {DLS} '14. {ACM} Press (2014)

\bibitem{Vitter1985}
Vitter, J.S.: Random sampling with a reservoir. {ACM} Transactions on
  Mathematical Software  \textbf{11}(1),  37--57 (Mar 1985)

\bibitem{Weyuker1982}
Weyuker, E.J.: On testing non-testable programs. The Computer Journal
  \textbf{25}(4),  465--470 (Nov 1982)

\end{thebibliography}

\end{document}